\documentclass[sn-mathphys]{sn-jnl}
\usepackage{graphicx}
\usepackage{amsmath}
\usepackage{amssymb}

\jyear{2021}%

\theoremstyle{thmstyleone}%
\theoremstyle{thmstyletwo}%

\theoremstyle{thmstylethree}%

\begin{document}

\title[Anisotropic star with a linear equation of state (EOS)]{Anisotropic star with a linear equation of state (EOS)}


\author[1]{\fnm{R.} \sur{Sharma}}\email{rsharma@associates.iucaa.in}
\equalcont{These authors contributed equally to this work.}

\author[2]{\fnm{B. S.} \sur{Ratanpal}}\email{bharatratanpal@gmail.com}
\equalcont{These authors contributed equally to this work.}

\author*[3]{\fnm{Rinkal} \sur{Patel}}\email{rinkalpatel22@gmail.com}
\equalcont{These authors contributed equally to this work.}

\affil[1]{\orgdiv{Department of Physics}, \orgname{Cooch Behar Panchanan Barma University},  \orgaddress{\street{CoochBehar},\city{WestBengal}, \postcode{736 101},  \country{India}}}

\affil[2]{\orgdiv{Department of Applied Mathematics}, \orgname{The Maharaja Sayajirao University of Baroda}, \orgaddress{\street{Faculty of Technology \& Engineering}, \city{Vadodara}, \postcode{390 001}, \state{Gujarat}, \country{India}}}

\affil*[1]{\orgdiv{Department of Applied Science \& Humanities}, \orgname{Parul University}, \orgaddress{\street{Limda}, \city{Vadodara}, \postcode{391 760}, \state{Gujarat}, \country{India}}}


\abstract{\noindent A family of solutions defining the interior of a static, spherically symmetric, compact anisotropic star is described by considering a new form of the equation of state (EOS). The analytic solution is derived by using the Finch and Skea ansatz for the metric potential $g_{rr},$ which has a clear geometric interpretation for the related background spacetime. The model parameters are fixed by smooth matching of the interior solution to the Schwarzschild exterior metric over the bounding surface of the compact star, together with the requirement that the radial pressure vanishes at the boundary. Data available for the pulsar $4U 18020 30$ has been utilized to analyze physical viability of the developed model. The model is shown to be stable.}


\maketitle

\section{Introduction}
\label{sec:1}
\noindent 
Since the groundbreaking work of Schwarzschild, generating an exact solution for a spherically symmetric perfect fluid distribution in general relativity has been subject of extensive study. Solutions to Einstein's field equations for geometrically meaningful spacetimes satisfying all the physical criteria are crucial in theoretical astrophysics. However, the non-linear nature of the Einstein field equations makes it difficult to find regular exact solutions fulfilling all the physical requirements. In addition, a feasible solution should also be able to describe realistic objects. 

In the high density regime of compact stars, linearity of the equation of state (EOS) of the matter composition appears to be a good approximation. \cite{nilsson2000general} studied static spherically symmetric perfect fluid stellar models with a linear barotropic EOS. \cite{ivanov2001relativistic} investigated relativistic static fluid spheres assuming a linear EOS. \cite{maharaj2006equation} developed new class of exact interior solutions to Einstein field equations and analyzed its physical behaviour. \cite{sharma2007class} obtained new exact solution to Einstein field equations making use of a linear EOS. New class of exact solutions to Einstein-Maxwell system was obtained by \cite{thirukkanesh2008charged}. \cite{maharaj2009generalized} also studied charged anisotropic matter distributions by assuming a linear EOS. \cite{varela2010charged} analyzed charged anisotropic star by considering linear as well as nonlinear EOS. \cite{maharaj2014some} developed a model for a quark star by considering a linear EOS. \cite{ngubelanga2015compact} obtained solutions to field equations in isotropic coordinates. \cite{harko2016exact} analyzed a power series solution for a stellar structure composed of an isotropic fluid which admits a linear barotropic or polytropic EOS. \cite{thomas2017anisotropic} studied anisotropic compact stars in paraboloidal spacetime with a linear EOS. Anisotropic compact stellar objects admitting
a linear EOS was studied by \cite{banerjee2018mathematical}. \cite{prasad2022anisotropic} presented a class of relativistic solutions to Einstein field equations for an anisotropic matter distribution utilizing the Buchdahl ansatz for the metric function $g_{rr}$. Recently, \cite{patel2023new} investigated a charged anisotropic stellar solution in paraboloidal spacetime using a linear EOS. All these studies are aimed at developing stellar models which are compatible with observational data.

While developing such models, one assumes a linear EOS the form $p_{r}= \alpha \rho -\beta,$ where $\rho$ is the density and $p_r$ is the radial pressure and $\alpha$ and $\beta$ are constants. Note that the linearity is in terms of density and not in terms of the radial variable $r$. This implies that $\alpha$ and $\beta$ might not be constants and could be the functions of the radial variable $r$ as well. Keeping this in mind in our work, to develop an anisotropic stellar model, we assume a linear EOS of state of the form $ p_{r}=\alpha\left( 1-\frac{r^2}{R^2}\right)\rho,$ where $0 <\alpha < 1$. This assumption allows us to generate a new class of exact solution to the Einstein field equations which is physical plausible.

The paper is organized as follow: Sec.~$2$ contains the Einstein field equations for a static spherically symmetric spacetime describing a star with energy momentum tensor for an anisotropic matter distribution. In Sec.~$3$, we solve the field equations by assuming a linear EOS as described earlier. In Sec.~$4$, the interior solution is matched to the Schwarzscchild exterior solution across the boundary $r=R$ of the stellar configuration. Physical plausibility of the model is discussed in Sec.~$5$. In Sec.~$6$, bounds on on the model parameters are  obtained and in Sec.~$7$ some concluding remarks are made.

\section{Field equations}
To develop the model of a static, spherically symmetric ansotropic star, we assume the spacetime metric in the form 
\begin{equation}\label{e1}
	ds^2 = e^{\nu(r)} dt^2-e^{\lambda(r)}dr^2 -r^2(d\theta^2 +sin^2 \theta d\phi^2).
\end{equation}
The energy-momentum tensor is assumed of the form
\begin{equation}\label{e2}
	T_{ij} = (\rho+p_{\perp})u_iu_j+pg_{ij}+\pi_{ij},
\end{equation}
where $ \rho $ and $p$ represent energy-density and isotropic pressure respectively and $ u_{i} $	is the unit $4$-velocity of fluid. The anisotropic stress-tensor $ \pi_{ij} $ is assumed to be of the form
\begin{equation}\label{e3}
	\pi_{ij} = \sqrt{3}S[C_{i}C_{j}-\frac{1}{3}(u_{i}u_{j}-g_{ij})],
\end{equation}
where $ S = S(r) $ denotes the magnitude of anisotropy and $ C^{i} = (0,-e^{\frac{\lambda}{2}},0,0) $ is a radially directed vector. We calculate the non-vanishing components of the	energy-momentum tensor as
\begin{equation}\label{5}
	T_{0}^{0} = \rho ,\;\;\;\;\;\;   T_{1}^{1}= -\left(p+\frac{2S}{\sqrt{3}}\right), \;\;\;\;\;\;\;\;     T_{2}^{2} = T_{3}^{3} = -\left(p-\frac{S}{\sqrt{3}}\right),
\end{equation}
and define the radial and tangential pressures as
\begin{equation}
	p_{r} = p +\frac{2S}{\sqrt{3}},     \;\;\;\;\;\;\;\;\;\;    p_{\perp} = p-\frac{S}{\sqrt{3}}.
\end{equation}
The magnitude of anisotropy obtained as
\begin{equation}\label{anisotropy}
	S = \frac{p_{r}-p_{\perp}}{\sqrt{3}}.
\end{equation}
The Einstein field equations, for the spacetime metric (\ref{e1}), together with the energy momentum tensor (\ref{e2}), leads to the following independent equations
\begin{equation}\label{rho}
	8\pi\rho = \frac{e^{-\lambda}\lambda'}{r}+\frac{1-e^{-\lambda}}{r^{2}},
\end{equation}
\begin{equation}\label{pr}
	8\pi p_{r} = \frac{e^{-\lambda}\nu'}{r}+\frac{e^{-\lambda}-1}{r^{2}},
\end{equation}
\begin{equation}\label{pp}
	8\pi p_{\perp} = e^{-\lambda} \left(\frac{\nu^{''}}{2} +\frac{\nu'^2}{4}-\frac{\nu' \lambda'}{4}+\frac{\nu'-\lambda'}{2r}\right),
\end{equation}
\begin{equation}\label{S}
	8\pi\sqrt{3}S = e^{-\lambda} \left(\frac{-\nu^{''}}{2} -\frac{\nu'^2}{4}+\frac{\nu' }{2r}+\frac{1}{r^2}-\frac{e^{\lambda}}{r^2}+\frac{\lambda'}{2r}+\frac{\nu' \lambda'}{4}\right).
\end{equation}

The technique to solve the system is discussed in the next section.

\section{Technique to generate new stellar solutions}
As we have three equations with five unknowns $ (\rho,p_{r},p_{\perp},e^{\lambda(r)},e^{\nu(r)})$, we can choose any two of them to close the system. This can be done in many different ways. For example, earlier ( \cite{sharma2013relativistic}-\cite{bhar2016new}) assumed specific forms of $\lambda(r)$ and $p_{r}$; \cite{bhar2015dark} assumed the density and radial pressure profiles; \cite{murad2015some} and \cite{thirukkanesh2018anisotropic} assumed particular form of $ \nu(r) $ together the the measure of anisotropy.  (\cite{sharma2007class}, \cite{thomas2017anisotropic}, \cite{bhar2016anisotropic}, \cite{sunzu2014charged}- \cite{ratanpal2023anisotropic}) assumed $ \lambda(r) $ and an EOS. In this paper, to develop a physically reasonable model of an anisotropic star, we
assume a linear EOS of the form
\begin{equation}\label{e4}
	8\pi p_{r} = \alpha (1-\frac{r^2}{R^2})\rho ,
\end{equation}
where $R$ is the radius of the star and $ 0 < \alpha < 1 $. Equation (\ref{e4}) guarantees that the radial pressure is positive at the center and vanishes at the boundary of the star.

We further use the Finch and Skea ansatz for the metric potential $ g_{rr} $ as
\begin{equation}\label{lambda}
	e^{\lambda(r)} = 1+\frac{r^2}{R^2},
\end{equation}
where $R$ is the curvature parameter. The ansatz (\ref{lambda}) has a geometric interpretation as can be found in reference \cite{tikekar2007relativistic}.

Combining equations (\ref{pr}) and (\ref{e4}), we obtain
\begin{equation}\label{e5}
	\nu^{'} = r\left[ e^{\lambda}\left(  \alpha \rho (1-\frac{r^2}{R^2})+\frac{1}{r^2}\right) -\frac{1}{r^2}\right].
\end{equation}
Integration of (\ref{e5}) yields
\begin{equation}\label{enu}
	e^{\nu} = CR^{4\alpha} (1+\frac{r^2}{R^2})^{2\alpha}\times exp\left( \frac{(1-\alpha)\frac{r^2}{R^2}}{2}-\frac{\alpha \frac{r^4}{R^4}}{4}\right),
\end{equation}
where $C$ is a constant of integration. Thus, the interior spacetime metric takes the form
{\footnotesize \begin{equation}\label{e6}
	ds^{2} = CR^{4\alpha} (1+\frac{r^2}{R^2})^{2\alpha} \times exp\left( \frac{(1-\alpha)\frac{r^2}{R^2}}{2}-\frac{\alpha \frac{r^4}{R^4}}{4}\right)dt^{2} - (1+\frac{r^2}{R^2})dr^{2} - r^2(d\theta^2+sin^2\theta d\phi^2),
\end{equation} }
which is non-singular at $r = 0$.

Making use of Eqs.~(\ref{e4}), (\ref{lambda}), (\ref{e5}) and (\ref{enu}), the system of equations (\ref{rho}-\ref{pp}) reduces to
\begin{equation}
	8\pi	\rho = \frac{3+\frac{r^2}{R^2}}{R^2(1+\frac{r^2}{R^2})^2},
\end{equation}	
\begin{equation}\label{e41}
	8\pi p_{r} = \frac{\alpha (1-\frac{r^2}{R^2})(3+\frac{r^2}{R^2})}{R^2(1+\frac{r^2}{R^2})^2},
\end{equation}
$	8\pi p_{\perp} =  $
{\footnotesize \begin{equation}
	\frac{12 \alpha + \alpha^2 \frac{r^{10}}{R^{10}}+2\alpha(2\alpha-1)\frac{r^8}{R^8}+(1-12\alpha-2\alpha^2)\frac{r^6}{R^6}-2(6\alpha^2+7\alpha-2)\frac{r^4}{R^4}+(3-16\alpha+9\alpha^2)\frac{r^2}{R^2}}{4R^2(1+\frac{r^2}{R^2})^3},
\end{equation}}
$ 8\pi\sqrt{3}S = $
{\footnotesize \begin{equation}
	\frac{-\frac{r^2}{R^2}\left( (3-20\alpha+9\alpha^2)-2\frac{r^2}{R^2}(6\alpha^2+\alpha-2)+\frac{r^4}{R^4}(1-8\alpha-2\alpha^2)+2\alpha\frac{r^6}{R^6}(2\alpha-1)+\frac{r^8}{R^8}\alpha^2\right) }{4R^2(1+\frac{r^2}{R^2})^3}.
\end{equation}}

\section{Exterior space-time and boundary conditions}

The model has three independent parameters, namely, $\alpha $, $C$, and $R$. Two of these constants can be evaluated by matching the interior spacetime metric (\ref{e6}) to the Schwarzschild exterior metric
\begin{equation}\label{e8}
	ds^{2} = \left(1-\frac{2M}{r}\right)dt^{2}-\left(1-\frac{2M}{r}\right)^{-1}dr^{2} - r^2(d\theta^2+sin^2\theta d\phi^2),
\end{equation} 
across the boundary $r = R$ of the star together with the condition that the radial pressure should vanish at the surface $(p_{r}(r = R) = 0)$. The process fixes the constants as
\begin{equation}\label{c}
	C = \frac{exp\left( \frac{3\alpha}{4}-\frac{1}{2}\right) }{2^{(2\alpha+1)} R^{4\alpha}},
\end{equation}
\begin{equation}\label{mass}
	M=\frac{ R}{4},
\end{equation}
The constant $C$ depends on $\alpha$, which remains as a free parameter.

\section{Physical conditions:}

For a physically acceptable stellar model, the following conditions should be satisfied ( \cite{finch1998review}, \cite{delgaty1998physical}):\\
$ (i) $\;\;\;\;   $ \rho(r) \ge 0 ,\;\;\;\;p_{r}(r)\ge 0 ,\;\;\;\;p_{\perp}(r)\ge 0 $   \;\;\;\;\;\;\;for   $ 0 \le r \le R  $.
\\$(ii) $ \;\;  $  \frac{d\rho}{dr}\le 0 ,\;\;\;\;  \frac{dp_{r}}{dr} \le 0  ,\;\;\;\;   \frac{dp_{\perp}}{dr} \le 0   $\;\;\;\;\;\; for  $ 0 \le r \le R  $
\\ $ (iii) $\;\;\;\; $0 \le \frac{dp_{r}}{d\rho} \le 1$ ,\;\;\;\; $0 \le \frac{dp_{\perp}}{d\rho} \le 1 $ \;\;\; for $ 0 \le r \le R  $  
\\  $(iv) $\;\;\;\;  $\rho-p_{r}-2p_{\perp}\ge 0 $ \;\;\;\; for $ 0 \le r \le R  $  
\\$(v) $\;\;\;\;  $\Gamma > \frac{4}{3}  $ \;\;\;\; for $ 0 \le r \le R. $ 

Using graphical method, we demonstrate that all of the above mentioned conditions are satisfied in this model.
The energy density in this model takes the form	
\begin{equation}
	8\pi	\rho = \frac{3+\frac{r^2}{R^2}}{R^2(1+\frac{r^2}{R^2})^2}.
\end{equation}
Thus, the central density takes the value
\begin{equation*}
	\rho(0) = \frac{3}{R^2}.
\end{equation*}
Obviously, we have $ \rho_{r=0} > 0 $ and $\rho_{r=R} > 0 $. The gradient of density is obtained as
\begin{equation}\label{e9}
	\frac{d\rho}{dr} = -\frac{2 \frac{r}{R} \left(5+\frac{r^2}{R^2}\right)}{R^3\left(r^2+R^2\right)^3},
\end{equation}
it can be shown from equation (\ref{e9}) that the density is a decreasing function of $r$. The condition ($ \frac{d\rho}{dr}_{r=0} < 0 $ and $ \frac{d\rho}{dr}_{r=R} < 0 $) puts the restriction on $ 0< \alpha < 1.$ The radial pressure 
\begin{equation}\label{e4*}
	8\pi p_{r} =  \frac{\alpha (1-\frac{r^2}{R^2})(3+\frac{r^2}{R^2})}{R^2(1+\frac{r^2}{R^2})^2}, 
\end{equation}
calculated the centre takes the form 
\begin{equation*}
	p_{r}(0)=\frac{3\alpha}{R^2}.
\end{equation*}
We note that the condition $ p_{r}(r=0) > 0 $  and $ p_{r}(r=R) > 0 $ are satisfies if $ 0 < \alpha < 1.$
Differentiating (\ref{e4*}) with respect to $r$, we obtain
\begin{equation}\label{e10}
	\frac{dp_r}{dr} = -\frac{16 \alpha  \frac{r}{R} }{R^3\left(r^2+R^2\right)^3},
\end{equation}
which is a decreasing function of $r$ provided $ 0 < \alpha < \frac{1}{4}$. The tangential pressure $p_{\perp} $ has the form
$	8\pi p_{\perp} =  $
{\footnotesize \begin{equation}
	\frac{12 \alpha + \alpha^2 \frac{r^{10}}{R^{10}}+2\alpha(2\alpha-1)\frac{r^8}{R^8}+(1-12\alpha-2\alpha^2)\frac{r^6}{R^6}-2(6\alpha^2+7\alpha-2)\frac{r^4}{R^4}+(3-16\alpha+9\alpha^2)\frac{r^2}{R^2}}{4R^2(1+\frac{r^2}{R^2})^3},
\end{equation}}
and its central value is
\begin{equation*}
	p_{\perp}(0) = \frac{3\alpha}{R^2}.
\end{equation*}
Thus, $ p_{\perp}(r=0) > 0 $. Also, the gradient of tangential pressure
$\frac{dp_{\perp}}{dr}= $
{\footnotesize \begin{equation*}
	\frac{\frac{r}{R}(9\alpha^2-52\alpha+3)+2(1+2\alpha-21\alpha^2)\frac{r^3}{R^3}+(6\alpha^2-22\alpha-1)\frac{r^5}{R^5}+8\alpha\frac{r^7}{R^7}+\alpha(9\alpha-2)\frac{r^9}{R^9}+2\alpha^2\frac{r^{11}}{R^{11}}}{2R^3\left(r^2+R^2\right)^4}.
\end{equation*}}
remains negative if $\alpha > \frac{1}{20}$. Thus, a more stringent bound on the parameter $\alpha$ is obtained as $ 0 < \alpha < \frac{1}{4} $. We also  note that the radial pressure and tangential pressure are equal at the centre implying regularity of the anisotropic factor. Fig.~(\ref{Figure:1}) shows the variation of density inside the star which decreases radially outward. Fig.~(\ref{Figure:2}) and Fig.~(\ref{Figure:3}) show variations of the radial and transverse pressures respectively. The two pressures are also decreasing functions of $r$.  Fig.~(\ref{Figure:4}) shows the anisotropy which is a decreasing throughout the distribution.

Let us now check whether the bound on $\alpha$ also satisfies the causality condition $0 < \frac{dp_{r}}{d\rho}< 1 $ and $0 <\frac{dp_{\perp}}{d\rho} < 1$. We have
\begin{equation*}
	\frac{dp_r}{d\rho}=\frac{8 \alpha  }{5+\frac{r^2}{R^2}},
\end{equation*}
$\frac{dp_{\perp}}{d\rho}= $
{\footnotesize \begin{equation*}
	\frac{\alpha(2-9\alpha)\frac{r^8}{R^8}+8\alpha(1-2\alpha)\frac{r^6}{R^6}+(1+22\alpha-6\alpha^2)\frac{r^4}{R^4}+2(21\alpha^2-2\alpha-1)\frac{r^2}{R^2}+(-3+52\alpha-9\alpha^2)}{4(1+\frac{r^2}{R^2})(5+\frac{r^2}{R^2})},
\end{equation*}}
The conditions $ 0\le \frac{dp_r}{d\rho}_{(r=0)}
\le 1 $ and $ 0 \le \frac{dp_r}{d\rho}_{(r=R)}
\le 1 $ are evidently satisfied at the centre as well as at the boundary.

The condition $ 0 \le \frac{dp_{\perp}}{d\rho}_{(r=0)}
\le 1$ and $ 0 \le \frac{dp_{\perp}}{d\rho}_{(r=R)}
\le 1$ are evidently satisfied at the centre as well as at the boundary provided $ \frac{1}{9} \left(26-\sqrt{649}\right)<\alpha <\frac{1}{9} \left(26-\sqrt{469}\right)$	and $ \frac{1}{20}<\alpha <\frac{13}{20}$. In Fig.(\ref{Figure:5}) and (\ref{Figure:6}), we show the variation of $\frac{dp_{r}}{d\rho}$ and $\frac{dp_{\perp}}{d\rho}$ against $r$. Both quantities satisfy the condition $ 0 <\frac{dp_{r}}{d\rho} < 1 $ and $ 0 < \frac{dp_{\perp}}{d\rho} < 1 $, indicating that the sound speed is less than the speed of light throughout the star. Table (\ref{tab:2}) shows the values of $ \frac{d\rho}{dr},$ $ \frac{dp_{r}}{dr}  $ and $ \frac{dp_{\perp}}{dr}$ at the center as well as the surface of the star. Table (\ref{tab:3}) shows the values of $ \frac{dp_{r}}{d\rho}  $ and $ \frac{dp_{\perp}}{d\rho}$ at the center as well as the surface of the star.

\subsection{Energy conditions:}
Conditions (i) and (ii) imply fulfillment of the weak and dominant energy	conditions. Condition (iv) ensures regular behaviour of the energy density. Now, we have
\begin{equation}
	(\rho - p_{r} - 2p_{\perp})_{(r=0)}=\frac{3 (1-3 \alpha )}{R^2},
\end{equation}
and
\begin{equation}
	(\rho - p_{r} - 2p_{\perp})_{(r=R)} = \frac{4 \alpha +1}{2 R^2}.
\end{equation}
In order to examine fulfillment of the strong energy condition, we evaluate $\rho - p_{r} - 2p_{\perp} $ at the centre and at the boundary of the star. It is observed that the bound on $ 0 < \alpha < \frac{1}{3} $ fulfills this condition. Fig.~(\ref{Figure:8}) indicates that the strong energy condition $\rho-p_{r}-2p_{\perp}>0$ is satisfied throughout the distribution within the bound of $\alpha $ where we have used the data obtained for the pulsar $4U 1820-30$. Table (\ref{tab:1}) shows the values of $ \rho-p_{r}-2p_{\perp}$ at the center as well as the surface of the star.

\subsection{Stability}
(i) Causality condition and method of cracking:
The stability of a stellar structure is critical in relativistic astrophysics. The causality criterion states that a physically plausible model's radial sound velocity $ v^{2}_{r} $ and transverse sound velocity  $ v^{2}_{\perp} $ must fall within the interval  $[0, 1]$. The expressions for the radial $ v^{2}_{r} $ and transverse $ v^{2}_{\perp} $ velocities of sound are obtained as
\begin{equation}
	v^{2}_{r} = \frac{p_{r}^{\prime}}{\rho^{\prime}} ,\;\;\;\ v^{2}_{\perp}=\frac{p_{\perp}^{\prime}}{\rho^{\prime}},
\end{equation}
\begin{equation}
	(v^{2}_{\perp}-v^{2}_{r})_{(r=0)} = \frac{1}{20} \left(-9 \alpha ^2+20 \alpha -3\right),
\end{equation}
\begin{equation}
	(v^{2}_{\perp}-v^{2}_{r})_{(r=R)} = \frac{1}{12} (4 \alpha -1),
\end{equation}
For $(v^{2}_{\perp}-v^{2}_{r})_{(r=0)} < 0$, we must have $ \left(-9 \alpha ^2+20 \alpha -3\right) < 0$ i.e., $0 < \alpha < 0.161777$. At the boundary of the star, we have
$(v^{2}_{\perp}-v^{2}_{r})_{(r=R)} < 0$. Thus, we must have $ (4 \alpha -1) < 0$ i.e., $0 <  \alpha < 0.240885.$ 

 \cite{herrera1992cracking} introduced the concept of ``cracking" to determine the stability of anisotropic matter distribution. \cite{abreu2007sound} showed that the region for which $ -1 \le v^{2}_{\perp}-v^{2}_{r} \le 0 $ are potentially stable and the region for which $ 0 \le v^{2}_{\perp}-v^{2}_{r} \le 1 $ are potentially unstable inside a stellar configuration. \cite{ratanpal2020cracking} analyzed the role of anisotropy in potentially stable or unstable regions based on the criteria put forward by Abreu {\em et al}. According to the theorem used by \cite{ratanpal2020cracking}, if $8\pi\sqrt{3}S = p_{r}-p_{\perp}$ is a decreasing function of $r$, then the stellar configuration is potentially stable. Table (\ref{tab:3}) shows that numerical values of the $ (v^{2}_{\perp}-v^{2}_{r}) $ at center as well as boundary of the star for the compact object $4U 1820-30$. Fig.(\ref{Figure:7}) shows that $ v^{2}_{\perp}-v^{2}_{r} < 0 $. Thus, the solution is potentially stable within the following bound: $0 < \alpha < 0.161777$.

(ii) Adiabatic index:\\	
 \cite{bondi1964contraction} showed that a Newtonian isotropic sphere will be in equilibrium if the adiabatic index $ (\Gamma) > 4/3$ which turns out to be true for a relativistic anisotropic fluid sphere as well. The adiabatic index $ \Gamma $ is given by
\begin{equation*}
	\Gamma_{r} = \frac{\rho+p_{r}}{p_{r}}\frac{dp_{r}}{d\rho},
\end{equation*}
\begin{equation}
	= \frac{8 \alpha  \frac{r^2}{R^2}-8 (\alpha +1) }{\left( \frac{r^4}{R^r}+4 \frac{r^2}{R^2}-5\right)  }.
\end{equation}
Within the prescribed bound of $ \alpha $, the profile of the adiabatic index ($ \Gamma _{r} $) is shown in Fig.(\ref{Figure:9}). The plot shows that the radial adiabatic index profile is a monotonic increasing function of $r$ and that $ \Gamma =\frac{\rho+p_{r}}{p_{r}}\frac{dp_{r}}{d\rho}>\frac{4}{3}$ everywhere inside the star thereby satisfying the stability requirement. Table \ref{tab:1} shows the value of $ \Gamma_{r}$ at the center of the star.

\subsection{Gravitational Redshift}
The redshift $ z = \sqrt{1/e^{\nu}} -1 $ must be a decreasing function of r and finite for $ 0\le z \le a $. For a relativistic star, it is expected that the redshift must decrease towards the boundary and be finite throughout the distribution. The value of redshift at origin is described in Table (\ref{tab:1})

As above all the conditions are satisfied in the range of $\alpha$ is $0.06< \alpha < 0.17.$	Therefore, our model is stable in the region $0.06< \alpha < 0.17.$

\begin{table}[h]
	\caption{Fulfillment of the strong energy condition and values of the gravitational redshift at the center as well as at the surface and adiabatic index at the surface where we have used the data for the pulsar $4U 1820-30$.}
	\label{tab:1}       
	\begin{tabular}{llllll} 
		\hline\noalign{\smallskip}
		\textbf{$\alpha$} & 
		{$ \mathbf{ \rho - p_{r} - 2p_{\perp}}_{(r=0)} $} & {$ \mathbf{\rho-p_{r}-2p_{\perp}}_{(r=R)} $} &    {$ \mathbf{ Z_{(r=0)}} $} &    {$ \mathbf{ Z_{(r=R)}} $} &  {$ \mathbf{\Gamma_{(r=0)}}$} 
		
		\\	&   \textbf{} & \textbf{} &    \textbf{(Redshift)} &    \textbf{(Redshift)} & \textbf{(Adiabatic }   \\
		& \textbf{}	&  \textbf{} & \textbf{} & \textbf{} &\textbf{ Index)}   \\
		\noalign{\smallskip}\hline\noalign{\smallskip}
		\textbf{$ 0.07 $} 	  & 861.84  & 232.73 &    0.312942   & 0 &1.71 \\
		\textbf{$ 0.08 $} 	  & 829.114   & 240.007 &  0.317126   & 0 & 1.72\\
		\textbf{$ 0.09 $} 	  & 796.38    & 247.28  &  0.321323   & 0   &1.74\\
		\textbf{$ 0.10 $}    & 763.65   & 254.55 &   0.325533   & 0 &1.76 \\
		\textbf{$ 0.11 $}    & 730.93  & 261.82 &  0.329757   & 0 & 1.77\\
		\textbf{$ 0.12 $} 	  & 698.201  & 269.09 &    0.333994   & 0 &1.79 \\
		\textbf{$ 0.13 $} 	  & 665.47   & 276.37 &    0.338245   & 0 & 1.8\\
		\textbf{$ 0.14 $} 	  & 632.74    & 283.644  & 0.342509   & 0   &1.82\\
		\textbf{$ 0.15 $}    & 600.017   & 290.917 &  0.346787   & 0 & 1.84 \\
		\textbf{$ 0.16 $}    & 567.28  & 298.19 &  0.351079   & 0 & 1.85\\

		\noalign{\smallskip}\hline
	\end{tabular} 
\end{table}
\begin{table}[h]
	\caption{Values of $ \frac{d\rho}{dr} $, $ \frac{dp_{r}}{dr} $ and  $ \frac{dp_{\perp}}{dr} $ at center as well as surface and at center as well as surface.}
	\label{tab:2}
	\begin{tabular}{lllllll}
		\hline\noalign{\smallskip}
		\textbf{$\alpha$} &  {$ \mathbf{\frac{d\rho}{dr}_{(r=0)}} $} & {$ \mathbf{\frac{d\rho}{dr}_{(r=R)}} $}  & {$ \mathbf{\frac{dp_{r}}{dr}_{(r=0)}} $}& {$ \mathbf{\frac{dp_{r}}{dr}_{(r=R)}} $} & {$ \mathbf{\frac{dp_{\perp}}{dr}_{(r=0)}} $} & {$ \mathbf{\frac{dp_{\perp}}{dr}_{(r=R)}} $}\\
		&   \textbf{} & \textbf{} & \textbf{}   \\
		\noalign{\smallskip}\hline\noalign{\smallskip}
		\textbf{$ 0.07 $} 	& 0  & -59.94  & 0&    -5.59   & 0 &-1.99 \\
		\textbf{$ 0.08 $} 	& 0  & -59.94  & 0 &   -6.39  & 0 & -2.99\\
		\textbf{$ 0.09 $} 	& 0  & -59.94    & 0  & -7.19 & 0   &-3.99\\
		\textbf{$ 0.10 $}  & 0  & -59.94   & 0 &  -7.99  & 0 & -4.99 \\
		\textbf{$ 0.11 $}  & 0  & -59.94  & 0 &  -8.79  & 0 & -5.99\\
		\textbf{$ 0.12 $} 	& 0  & -59.94  & 0 &    -9.59   & 0 &-6.99 \\
		\textbf{$ 0.13 $} 	& 0  & -59.94   & 0 &   -10.38  & 0 & -7.99\\
		\textbf{$ 0.14 $} 	& 0  & -59.94   & 0  & -11.18 & 0   &-8.99\\
		\textbf{$ 0.15 $}   & 0 & -59.94   & 0 &  -11.98  & 0 &-9.99 \\
		\textbf{$ 0.16 $}  & 0  & -59.94  & 0&  -12.78  & 0 & -10.98\\
		
		\noalign{\smallskip}\hline
	\end{tabular} 
\end{table}

\begin{table}[h]
	\caption{Values of $ \frac{dp_{r}}{d\rho} $  and $ \frac{dp_{\perp}}{d\rho} $ at the center as well as at the surface and at center.}
	\label{tab:3}
	\begin{tabular}{lllllll}
		\hline\noalign{\smallskip}
		\textbf{$\alpha$} &  {$ \mathbf{\frac{dp_{r}}{d\rho}_{(r=0)}} $} & {$ \mathbf{\frac{dp_{\perp}}{d\rho}_{(r=0)}} $}  & {$ \mathbf{\frac{dp_{r}}{d\rho}_{(r=R)}} $} & {$ \mathbf{\frac{dp_{\perp}}{d\rho}_{(r=R)}} $} & {$ \mathbf{(\nu^{2}_{t}-\nu^{2}_{r})_{(r=0)}} $} & {$ \mathbf{(\nu^{2}_{t}-\nu^{2}_{r})_{(r=R)}} $}\\
		&   \textbf{} & \textbf{} & \textbf{}   \\
		\noalign{\smallskip}\hline\noalign{\smallskip}
		\textbf{$ 0.07 $} 	  & 0.112  & 0.029 &    0.093   & 0.033 &-0.082 &-0.06 \\
		\textbf{$ 0.08 $} 	  & 0.128   & 0.055 &   0.106  & 0.055 & -0.072&-0.056\\
		\textbf{$ 0.09 $} 	  & 0.144   & 0.08  & 0.12 & 0.066   &-0.063& -0.053\\
		\textbf{$ 0.10 $}    & 0.16   & 0.105 &  0.13  & 0.083 &-0.054&-0.05 \\
		\textbf{$ 0.11 $}    & 0.176  & 0.130 &  0.146  & 0.1 & -0.045&-0.046\\
		\textbf{$ 0.12 $} 	  & 0.192  & 0.15 &    0.16   & 0.11 &-0.036&-0.043 \\
		\textbf{$ 0.13 $} 	  & 0.208   & 0.180 &   0.173  & 0.133 & -0.027&-0.04\\
		\textbf{$ 0.14 $} 	  & 0.224   & 0.205  & 0.186 & 0.15  &-0.018&-0.036\\
		\textbf{$ 0.15 $}    & 0.24   & 0.229 &  0.2 & 0.16 &-0.010&-0.033 \\
		\textbf{$ 0.16 $}    & 0.256  & 0.254 &  0.213  & 0.183 & -0.0015&-0.03\\
		
		\noalign{\smallskip}\hline
		
	\end{tabular} 
	
\end{table}
\section{Conclusions}
\begin{figure}[H]\centering
	\includegraphics[scale = 1]{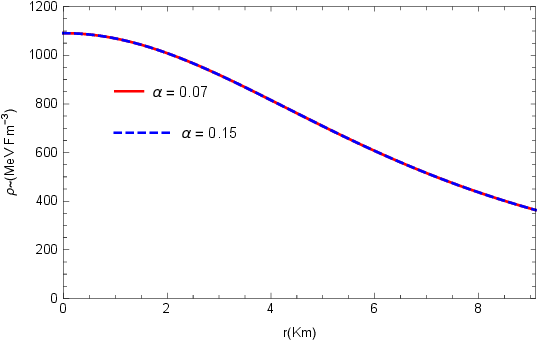} 
	\caption{Variation of density against radius $r$.}
	\label{Figure:1}
\end{figure}

\begin{figure}[H]\centering
	\includegraphics[scale = 1]{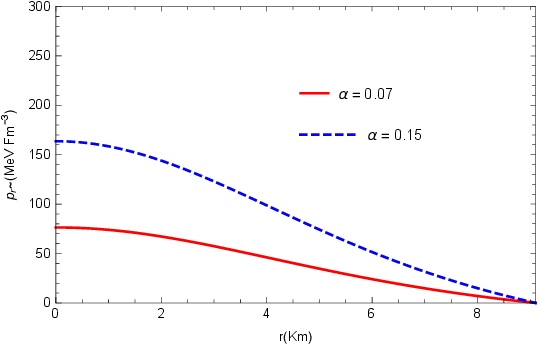}
	\caption{Variation of radial pressure against radius $r$.}
	\label{Figure:2}
\end{figure}

\begin{figure}[H]\centering
	\includegraphics[scale = 1]{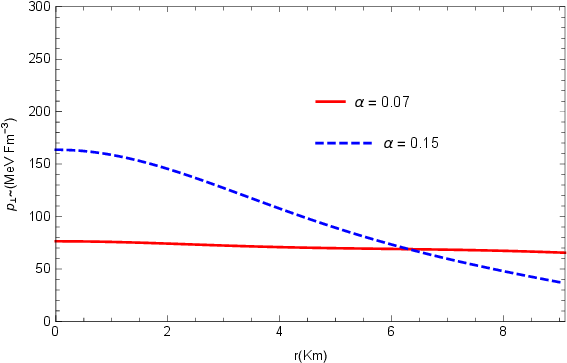}
	\caption{Variation of transverse pressure against radius $r$ 
	}
	\label{Figure:3}
\end{figure}

\begin{figure}[H]\centering
	\includegraphics[scale = 1]{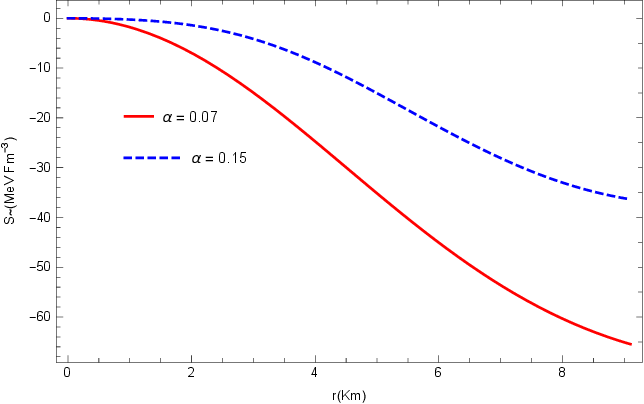}
	\caption{Variation of anisotropy against radius $r$. 
	}
	\label{Figure:4}
\end{figure}

\begin{figure}[H]\centering
	\includegraphics[scale = 1]{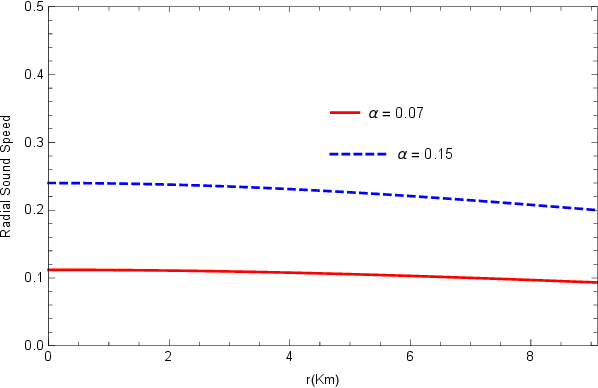}
	\caption{Variation of $ \frac{dp_r}{d\rho} $ against radius $r$. 
	}
	\label{Figure:5}
\end{figure}

\begin{figure}[H]\centering
	\includegraphics[scale=1]{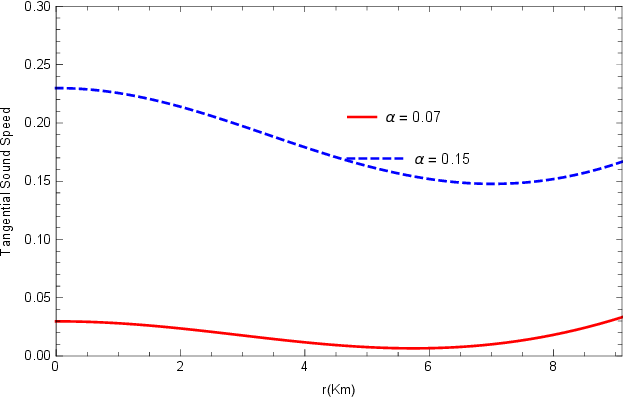}
	\caption{Variation of $ \frac{dp_\perp}{d\rho} $ against radius $r$.
	}
	\label{Figure:6}
\end{figure}

\begin{figure}[H]\centering
	\includegraphics[scale = 1]{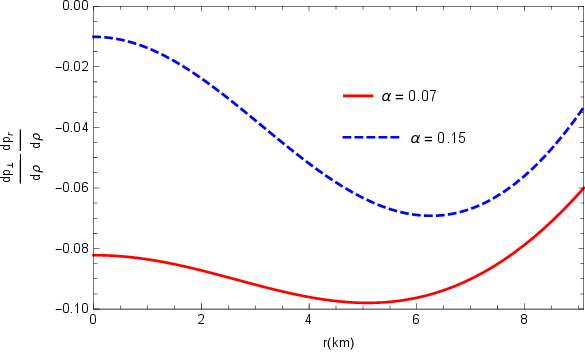} 
	\caption{The causality
		condition plotted with respect to the radial coordinate $r$.}
	\label{Figure:7}
\end{figure}

\begin{figure}[H]\centering
	\includegraphics[scale = 1]{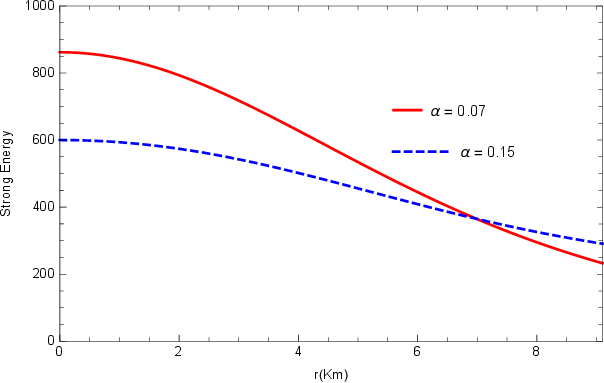}
	\caption{Fulfillment of the strong energy condition against radial variable $r$. 
	}
	\label{Figure:8}
\end{figure}

\begin{figure}[H]\centering
	\includegraphics[scale = 1]{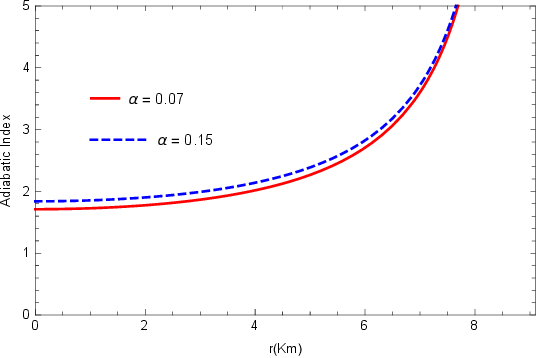}
	\caption{Variation of adiabatic index against radius $r$. 
	}
	\label{Figure:9}
\end{figure}

\begin{figure}[H]\centering
	\includegraphics[scale = 1]{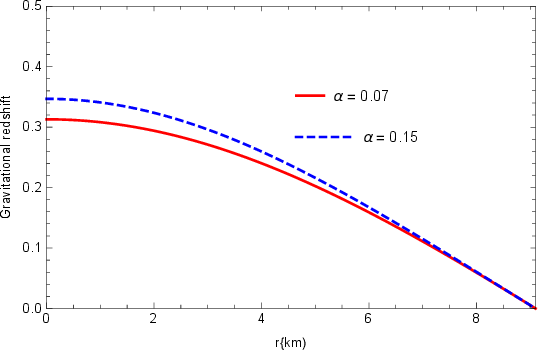}
	\caption{Variation of Gravitational redshift. 
	}
	\label{Figure:10}
\end{figure}

\begin{figure}[H]\centering
	\includegraphics[scale = 1]{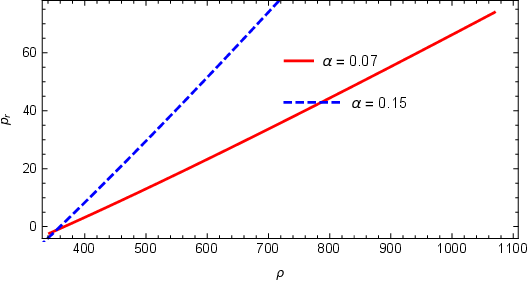}
	\caption{The relation between radial pressure $p_{r}$ and density $\rho$.}
	\label{Figure:11}
\end{figure}

In the present work, we solved Einstein's field equations defining a spherically symmetric anisotropic matter by assuming the Finch and Skea ansatz and considering a linear equation of state of the form $ p_{r}=\alpha\left( 1-\frac{r^2}{R^2}\right)\rho,$ where $0<\alpha < 1$. Physical grounds have been used to get bounds on the model parameters, and it has been demonstrated that the model is stable for $ 0.06 < \alpha < 0.17.$   All the physical quantities are regular and well-behaved throughout the stellar interior for the star 4U1820-30 with radius R = 9.1 km and mass $ M =  1.58M_\odot $. In Fig.(\ref{Figure:1}), Fig.(\ref{Figure:2}) and Fig.(\ref{Figure:3}), we examine the physical matter variables $\rho$, $p_{r}$, $ p_{\perp}$ graphically. In addition, the anisotropy for the model is shown to decreasing, as seen in Fig.(\ref{Figure:4}). Fig.(\ref{Figure:5}) and Fig.(\ref{Figure:6}) shows both the radial and tangential square of sound speed. Fig.(\ref{Figure:7}) shows that $ v^{2}_{\perp}-v^{2}_{r} < 0 $ throughout the star. The energy criterion is met within the stellar structure.  Since positive density and pressure are bound to be $\ge 0$, we investigate the profile of the SEC $(\rho-p_{r}-2p_{\perp})$ graphically to confirm the stability in Fig.(\ref{Figure:8}), and it is found to be satisfied for our model. We examined the adiabatic index, which is greater than $\frac{4}{3}$ across the structure (see Fig.(\ref{Figure:9})). It can be seen that the redshift maximizes at the centre shown in Fig.(\ref{Figure:10}).  We have shown that the model admits an equation of state which is Linear in nature which is shown with graphical representation in Fig.(\ref{Figure:11}). So the presented model satisfies all the physical criteria of a physically well-behaved compact object in the region $ 0.06 < \alpha < 0.17.$



\section*{Acknowledgement}
BSR and RP are thankful to IUCAA Pune for the facilities and hospitality provided to them where part of the work was carried out. RS gratefully acknowledges support from the Inter-University Centre for Astronomy and Astrophysics (IUCAA), Pune, India, under its visiting Research Associateship Programme. 


\end{document}